\documentstyle[aps,preprint,epsfig,amssymb]{revtex}
\tightenlines

\begin{document}

\title{Energy spectra of quantum rings}

\author{A. Fuhrer$^1$, S. L\"uscher$^1$, T. Ihn$^1$, T. 
Heinzel$^{1,2}$, K. Ensslin$^1$,W. Wegscheider$^3$, and M. Bichler$^4$}

\address{$^1$ Solid State Physics Laboratory, ETH Z\"urich, 8093 
Z\"urich, Switzerland,\\
$^2$ Angewandte und Experimentelle Physik, Universit\"at Regensburg, 
93040 Regensburg, Germany,\\
$^3$ Walter Schottky Institut, TU M\"unchen, 85748 Garching, 
Germany,\\
$^4$ Fakult\"at f\"ur Physik, Universit\"at Freiburg i. Br., 79104 
Freiburg, Germany}

\date{\today}

\maketitle

\begin{abstract}
Ring geometries have fascinated experimental and theoretical 
physicists over many years. Open rings connected to leads allow the 
observation of the Aharonov-Bohm effect \cite{1.}, a paradigm of quantum 
mechanical phase coherence \cite{2.,3.}. The phase coherence of transport 
through a quantum dot embedded in one arm of an open ring has been 
demonstrated \cite{4.}. The energy spectrum of closed rings \cite{5.} has only 
recently been analysed by optical experiments \cite{6.,7.} and is the basis 
for the prediction of persistent currents \cite{8.} and related experiments 
\cite{9.,10.,11.}. Here we report magnetotransport experiments on a ring-shaped 
semiconductor quantum dot in the Coulomb blockade  regime \cite{12.}. The 
measurements allow us to extract the discrete energy levels of a 
realistic ring, which are found to agree well with theoretical 
expectations. Such an agreement, so far only found for few-electron 
quantum dots, is here extended to a many-electron system \cite{13.}. In a 
semiclassical language our results indicate that electron motion is 
governed by regular rather than chaotic motion, an unexplored regime 
in many-electron quantum dots.
\end{abstract}

\newpage
Quantum ring samples have been fabricated on AlGaAs-GaAs 
heterostructures containing a two-dimensional electron gas (2DEG) 
with density $\mathrm{5\times 10^{15} m^{-2}}$ and mobility $\mathrm{900'000 
cm^2/Vs}$ at $\mathrm{T=4.2 K}$ only 
34 nm below the sample surface. The surface of the heterostructure 
has been locally oxidized by applying a voltage between the 
conductive tip of an atomic force microscope (AFM) and the 2DEG \cite{14.}. 
The electron gas is depleted below the oxidized regions, which was 
used in other studies for defining high quality quantum dots \cite{15.}. The 
details of the fabrication process which is crucial for the 
high-electronic quality of the quantum ring, are described in Ref. 
\cite{16.}. Figure 1(a) shows an AFM image (taken with an unbiased tip 
directly after the oxidation process) of the oxide lines defining the 
quantum ring. The width of the quantum point contacts connecting the 
ring to source (drain) is controlled by voltages applied to the 
lateral gate electrodes qpc1a and b (qpc2a and b). The number of 
electrons in the ring can be tuned via the lateral plunger gates pg1 
and 2. Shape deformations due to applied in-plane gate voltages are 
known to be relatively weak \cite{15.,16.}. The schematic in Fig. 1(b) shows 
the dimensions of the quantum ring.
After the oxidation step the sample has been covered with a metallic 
top gate electrode. With the combination of in-plane and top gate 
electrodes the quantum ring can be tuned into the Coulomb blockade 
regime with the single-particle level spacing being much larger than 
the thermal energy kT. 
Figure 2(b) presents a colour plot of the current through the 
quantum ring as a function of plunger-gate voltage and magnetic field 
B (applied normal to the 2DEG plane). This measurement was performed 
at a source-drain voltage $\mathrm{V_{SD} = 20 \mu V}$ and at a temperature of 100mK 
in a dilution refrigerator. In Fig. 2(a) the Coulomb blockade 
oscillations have been extracted along the horizontal dashed line in 
Fig. 2(b), i.e. at constant B=92 mT. From corresponding measurements 
of the Coulomb blockade  diamonds we determine a charging energy 
$\mathrm{E_{C}=e^2/C_{\Sigma} \approx 190\mu eV}$. The observed discrete level spacings after 
subtraction of Ec within the constant interaction model \cite{12.} can be as 
large as $\mathrm{\Delta \approx 180\mu eV}$ (see below).  From a simple capacitance model and 
the ring geometry we estimate that about 200 electrons are 
distributed on 2-3 radial subbands. In Fig. 2(b) the position as well 
as the amplitude of the Coulomb blockade peaks oscillate as a 
function magnetic field with a period of $\mathrm{\Delta B=75mT}$ (horizontal white 
lines) which is exactly the Aharonov-Bohm period for this ring. In 
fact, by opening the point contacts (not shown) we find the 
well-known Aharonov-Bohm oscillations in the conductance with the 
same $\mathrm{\Delta B}$ \cite{2.,3.,17.,18.}. The oscillations are visualized along a line of 
constant gate voltage in Fig. 2(c) (Fig. 2(b), vertical dashed line).

What determines the magnetic field dependence of positions and 
amplitudes of the Coulomb blockade  resonances? We start the 
discussion from the energy spectrum of an infinitely thin perfect 
ring of radius $r_{0}$ enclosing \textit{m} flux quanta: \cite{8.}

\[
E_{m,l} = \frac{\hbar^2}{2 m^\star r_{0}^2}(m+l)^2
\]

Here $m^\star$ is the mass of the particle, \textit{l} is the angular momentum quantum 
number. For a given angular momentum state the energies as a function 
of magnetic field (or flux \textit{m}) lie on a parabola with its apex at \textit{m} = 
-\textit{l}, as depicted in Figure 3(a). According to this simple picture a 
single Coulomb blockade  peak should oscillate as a function of B 
along a zig-zag line (blue curve) with the Aharonov-Bohm period 
$\mathrm{\Delta B}$. 
Comparison with the measurement in Fig. 2(b) shows that indeed some 
peaks move along a zig-zag line, but others show barely any 
B-dependence, a behaviour which will be discussed later.

We first take a closer look at the h/e-periodic modulation of the 
Coulomb blockade peak amplitude. A peak follows a line of constant 
electron number (blue line in Fig. 2(a)). The current is successively 
carried by states (\textit{l}), (-\textit{l}-1),(\textit{l}-1),(-\textit{l}-2),(\textit{l}-2),Éwhen B is 
increased from zero, i.e. a change in state occurs every half flux 
quantum. The amplitude of Coulomb blockade peaks is determined by the 
wave function overlap between the confined states in the ring and the 
extended states in source and drain \cite{12.}. However, the wave functions 
in an infinitely thin perfect ring are independent of magnetic field 
and given by

\[
\Psi_{l}(\Phi) = \frac{1}{\sqrt{2 \pi}} e^{i l \Phi}
\]

where $\Phi$ is the azimutal coordinate in the ring. The (lateral) overlap 
(proportional to $|\Psi|^2$) with source and drain is the same for all 
states. This model does therefore not predict the observed 
h/e-periodic modulation of the peak amplitude.

A more realistic but still analytically soluble model takes the 
finite extent of the wave functions in radial direction into account 
\cite{19.} , which leads to multiple radial channels indexed by the quantum 
number n. Using this model and analysing the Coulomb diamonds in 
detail we obtain agreement with the previous estimate of about 2-3 
radial modes and about 200 electrons in the ring. For small (large) 
n, the occupied states at the Fermi level have a larger (smaller) 
angular momentum quantum number \textit{l} at B=0 and consequently display a 
stronger (weaker) magnetic field dispersion. This model predicts that 
the exponential decay of the wave functions in radial direction 
depends on the value of \textit{l} but is relatively insensitive to the value 
of \textit{m} (for small \textit{m}). Since all states move in zig-zag lines with an 
h/e periodicity in magnetic field, crossings of states with different 
angular momenta \textit{l} may lead to a different wave function overlap and 
therefore to a modulation of the corresponding Coulomb peak amplitude.

The angular uniformity of the probability density in a perfect ring 
stems from the cylindrical symmetry which, for the real sample, will 
be broken by the pure presence of source and drain, by dopants and by 
the limits of the fabrication procedure.  This leads to pinning of 
the wave function and therefore to a distinct amplitude of the 
probability density at source and drain. The perturbation will become 
especially important at the degeneracy points of levels where it 
leads to anti-crossing behaviour \cite{20.}. In the simplest case the 
probability density changes from a uniform to a sinusoidal 
angle-dependence at the degeneracy points. In this picture the 
AB-periodic oscillation of the amplitude along a single Coulomb peak 
can be understood in terms of changing contributions of single 
particle levels to the current carrying state. 

We now turn to the analysis of the experimental Coulomb blockade  
peak positions.  They are obtained from measurements like the ones 
shown in Fig. 2(b) by converting the gate-voltage axis into an energy 
scale using the appropriate lever arm as determined from the analysis 
of the Coulomb blockade diamonds \cite{21.}. A constant charging energy of 
$\mathrm{190\mu eV}$ is subtracted from the position of a Coulomb maximum \cite{12.,15.} and 
the resulting energies are plotted in Fig. 3(c) as a function of 
magnetic field. Clearly many of the peaks move in pairs (see e.g. the 
black-purple, the blue-red or the green-purple pair), previously 
identified as spin pairs \cite{15.}. The exchange related spin splitting 
energy with a value of $\mathrm{\approx 20 \mu eV}$ for these peaks is on average smaller 
than the discrete energy level spacing $\mathrm{\Delta}$. Electrons therefore 
successively populate these orbital states with spin-up and spin-down 
electrons. For other peaks (green, yellow, red) spin pairing is not 
clearly observed. As depicted in Fig. 3(c) the orbital states move up 
and down in magnetic field with the Aharonov-Bohm period $\mathrm{\Delta B}$. In this 
respect our experiments show the long-predicted energy spectrum 
characteristic for quantum rings \cite{5.}.

Let us look at the states in Fig. 3(c) that have a very small 
dispersion in the magnetic field. In the framework of the model of 
Ref. \cite{19.} such 'flat' states can occur at the onset of the occupation 
of the next higher radial channel. However, in the experiment we 
observe such states over wide ranges of gate voltages and an extended 
model is necessary. In Fig. 3(b) we show a calculation with 
ring-parameters typical for our dot. The spectrum is obtained from 
the diagonalisation of a truncated Hamiltonian matrix expressed in 
the eigenstate-basis of Ref. \cite{19.}, but including diagonal and 
off-diagonal elements given by a symmetry breaking potential \cite{22.}. Two 
radial modes are taken into account leading to two sets of parabolic 
energy dispersions as a function of magnetic field in the unperturbed 
system. The perturbation mixes states of positive and negative 
angular momenta, which leads in some cases to eigenenergies that 
barely depend on magnetic field. Such states can intersect the 
diamonds formed by strongly oscillating levels in close resemblance 
to the experimental findings. The model obviously can only give the 
general tendency of the experimental spectra. Nevertheless, the 
dominant deviations from the perfect ring spectrum can be understood 
as the result of a symmetry breaking potential perturbation.

We estimate the contribution of a particular strongly oscillating 
state to the persistent current \cite{8.} from the experimental dispersion 
(i.e. black dots in Fig. 3(c)) and obtain a value of 5nA. If we 
assume that currents of all the lower lying states sum up to zero 
this current is also an estimate of the total persistent current in 
the ring and the value is consistent with previous magnetisation 
measurements \cite{9.,10.,11.}.

For quantum dots containing a small number of electrons, the shell 
structure of the level occupancy can clearly be detected because of 
the dominating cylindrical symmetry \cite{13.}. Quantum dots containing many 
electrons are usually described in the context of an underlying 
classically chaotic geometry, since small perturbations of parameter 
space such as potential shape or magnetic field can induce parametric 
fluctuations in the energy levels and consequently in the Coulomb 
peak positions. Several theoretical and experimental publications 
address the question whether the spectra of such many-electron 
quantum dots can be adequately described by Random Matrix Theory \cite{24.}. 
Our quantum ring represents a many-electron Coulomb blockaded system 
with regular geometry. This together with the small number of radial 
modes is the reason why we can qualitatively understand the principal 
features of the observed energy spectrum without the need of a 
statistical analysis. The source of this striking observation lies in 
the circular geometry of our ring.

In order to support this view we have fabricated a square shaped 
quantum dot with a circular antidot in the centre. This system is 
considered a Sinai billiard and is one of the theoretically best 
studied examples of a classically chaotic system. Fig. 4 shows the 
evolution of the conductance as a function of plunger gate and 
magnetic field of this system presented in a way comparable to Fig. 
2(b) for the quantum ring. Around B=0 the Coulomb peak maxima 
fluctuate irregularly as a function of magnetic field with an \textit{average} 
period compatible with an Aharonov-Bohm-type argument. As the 
magnetic field is increased to a value, where the classical cyclotron 
diameter matches the antidotÕs circumference, a well pronounced 
B-periodic behaviour of the Coulomb peak maxima in amplitude and 
position is recovered. Similar as in antidot lattices \cite{25.} the magnetic 
field is expected to induce regular parts in the predominately 
chaotic phase space existing at B=0. 

The detailed analysis of quantum rings demonstrates that even in 
many-electron Coulomb blockaded systems a detailed understanding of 
the energy spectrum can be obtained. With advanced fabrication 
techniques at hand this opens the path to the understanding of more 
complex and multiply connected structures on a quantum mechanical 
level. Electron-electron interactions beyond the constant interaction 
model are believed to play a minor role in our quantum ring. One 
indication is the observation of spin pairs and relatively small spin 
splitting. Once ring structures with only one radial mode occupied 
are available such quantum rings could be used to investigate spin 
effects \cite{26.} or even Luttinger liquid behaviour in a circular 1D system 
with periodic boundary conditions. One can also envision searching 
for persistent current effects in the transport signatures of a 
quantum ring now that the energy spectrum is experimentally 
accessible.
\vskip 1cm
We thank M. B\"uttiker and D. Loss for valuable discussions. Financial 
support from the Swiss Science Foundation (Schweizerischer 
Nationalfonds) is gratefully acknowledged.

\newpage
\begin{figure}
\noindent\epsfig{file=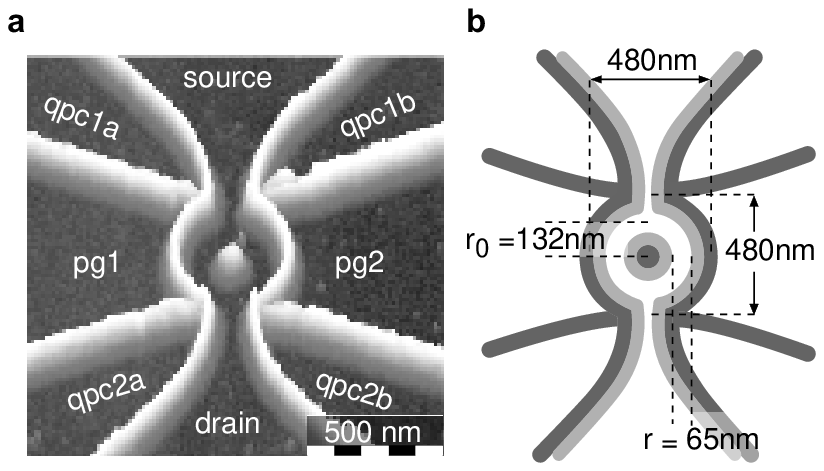,width=0.6\linewidth,angle=0}
\vskip 1cm 
\caption{Sample layout. {\sffamily\textbf{a}}, Micrograph of the quantum ring taken with 
the unbiased AFM-tip after writing the structure. The oxide lines 
(bright regions) deplete the 2DEG 34nm below the surface separating 
the sample into several conductive (dark) regions. The current is 
passed from source to drain. The in-plane gates (qpc1a, qpc1b, qpc2a, 
qpc2b, pg1 and pg2) are used to tune the point contacts and two arms 
of the ring. {\sffamily\textbf{b}}, Schematic sketch of the ring. The dark curves 
represent the oxide lines. From transmission measurements of the 
point contacts at source and drain we estimate the depletion length 
to be about 50nm which results in an estimated channel width of 
$\Delta r \approx$ 65nm. The average radius of the ring is $\mathrm{r_{0} = 132nm}$.}
\label{fig1}
\end{figure}

\newpage
\begin{figure}
\noindent\epsfig{file=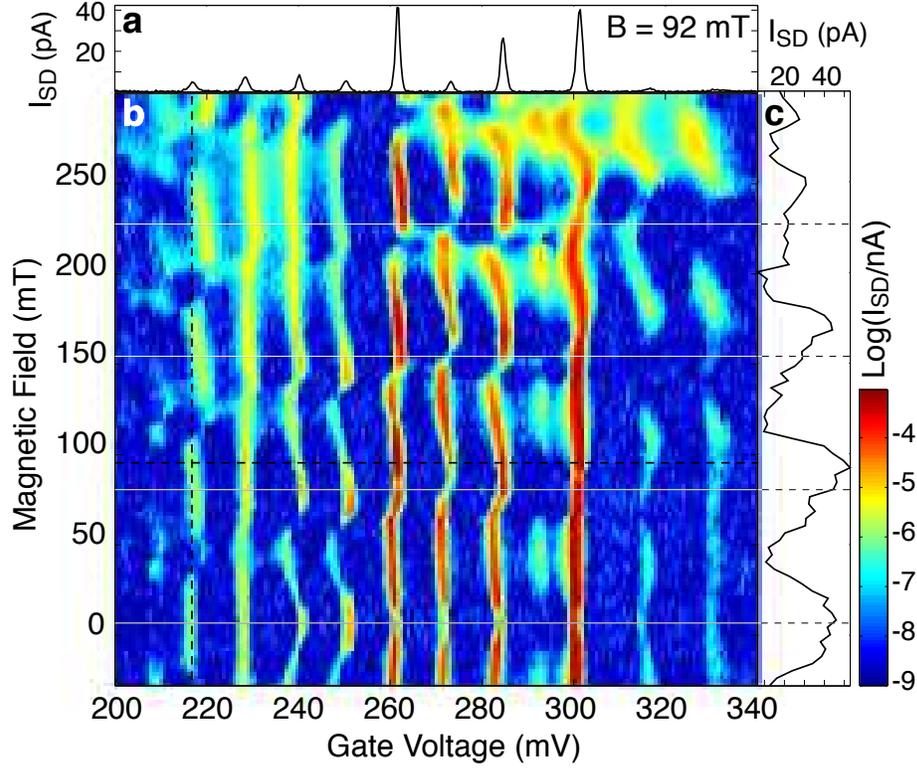,width=0.8\linewidth,angle=0}
\vskip 1cm 
\caption{The addition spectrum. {\sffamily\textbf{a}}, Measurement of Coulomb blockade 
resonances at fixed magnetic field. The current is measured as a 
function of a voltage applied to both plunger gates (pg1 and 2) 
simultaneously. {\sffamily\textbf{b}}, The evolution of such sweeps with magnetic field 
results in the addition spectrum shown in colour. The regions of high 
current (yellow/red) mark configurations in which a bound state in 
the ring aligns with the Fermi level in source and drain. The 
Aharonov-Bohm period expected from the ring geometry is indicated by 
the thin white horizontal lines. {\sffamily\textbf{c}}, Magnetic field sweep for constant 
gate voltage Vpg = 218mV (dashed line in the colour plot). While this 
peak shows a maximum in amplitude for B=0 other peaks (Vpg = 270mV) 
display a minimum in amplitude.}
\label{fig2}
\end{figure}

\newpage
\begin{figure}
\noindent\epsfig{file=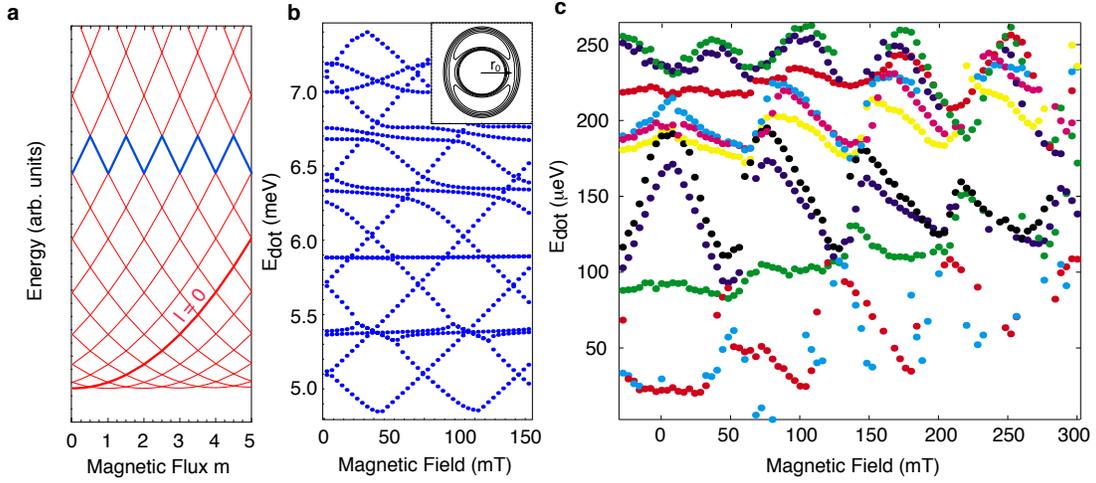,width=1\linewidth,angle=0}
\vskip 1cm 
\caption{Reconstruction of the energy spectrum. {\sffamily\textbf{a}}, Theoretical 
spectrum of a single mode ring. The parabolas with constant \textit{l} (bold 
red line) have a minimum at \textit{l = m}. The blue zig-zag line corresponds 
to a Coulomb peak after the charging energy has been subtracted in 
the constant interaction model. {\sffamily\textbf{b}}, Calculated spectrum with ring 
parameters typical for our dot. We assume a slightly asymetric 
potential shown in the inset, which mixes states of positive and 
negative angular momenta. This leads in some cases to eigenenergies 
that barely depend on magnetic field. {\sffamily\textbf{c}}, Reconstruction of the energy 
spectrum of the ring from the data shown in Figure 2. The plunger 
gate voltage was converted into dot energy using measurements of the 
Coulomb diamonds and a constant charging energy of 190$\mu$eV was 
subtracted.}
\label{fig3}
\end{figure}

\newpage
\begin{figure}
\noindent\epsfig{file=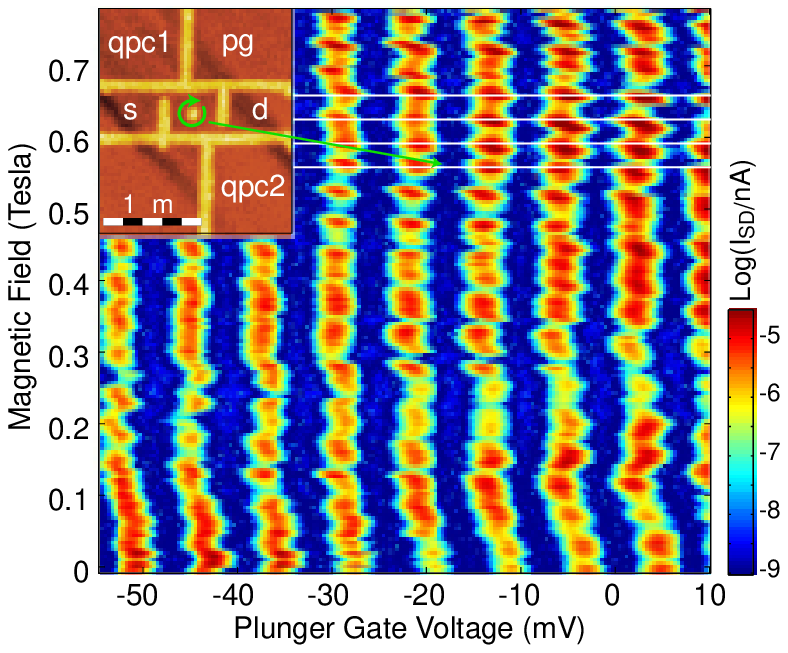,width=0.8\linewidth,angle=0}
\vskip 1cm 
\caption{Addition spectrum of a Sinai billiard. The inset shows an 
AFM-micrograph of the square shaped quantum dot with a circular 
antidot in the centre. The colour plot is a similar measurement to 
the one in Figure 2(b) and shows the evolution of the conductance as 
a function of plunger gate voltage and magnetic field. Around B=0 the 
Coulomb maxima fluctuate irregularly as the magnetic field is 
changed. As the magnetic field is increased to a value, where the 
classical cyclotron orbit matches the antidotÕs circumference 
(indicated by the green circle and the arrow in the inset), a well 
pronounced h/e-periodic behaviour (white lines) in amplitude and 
position of the Coulomb peaks is recovered, indicating quenching of 
the chaotic behaviour.}
\label{fig4}
\end{figure}

\begin{thebibliography}{99}
\bibitem{22.} The angle independent part $\mathrm{\omega_{0}}$ of Ref. 
\cite{19.} is replaced by $\mathrm{\omega_{0} (1+\epsilon 
\cos{2\phi})}$ following the idea of shape deformation in Ref. \cite{23.}. This 
breaks the symmetry leading to inter- as well as intra-subband 
coupling. The values used for the calculation were $\mathrm{\omega_{0} = 
2.73\times 10^{12} s^{-1}}$, 
$\mathrm{r_{0}=132nm}$ and $\mathrm{\eta=0.06}$.
\end{thebibliography}
\end{document}